\documentclass[a4paper,12pt]{article}

\usepackage{graphics, epsfig}

\begin{document}

\author{C. Barrab\`es \thanks{E-mail : barrabes@lmpt.univ-tours.fr}\\
\small Laboratoire de Math\'ematiques et Physique Th\'eorique,\\
\small CNRS/UMR 6083, Universit\'e F. Rabelais, 37200 TOURS,
France
\\\small and \\P. A. Hogan \thanks{E-mail : peter.hogan@ucd.ie} \\
\small School of Physics, \\ \small University College Dublin, Belfield, Dublin 4, Ireland}

\title{On The Interaction of Gravitational Waves with Magnetic and Electric Fields}
\date{}
\maketitle

\begin{abstract}
The existence of large--scale magnetic fields in the universe has led to the observation that 
if gravitational waves propagating in a cosmological environment encounter even a small magnetic field 
then electromagnetic radiation is produced. To study this phenomenon in more detail we take it out of the 
cosmological context and at the same time simplify the gravitational radiation to impulsive waves. Specifically, 
to illustrate our findings, we describe the following three physical situations: (1) a cylindrical impulsive 
gravitational wave propagating into a universe with a magnetic field, (2) an axially symmetric impulsive 
gravitational wave propagating into a universe with an electric field and (3) a `spherical' impulsive gravitational 
wave propagating into a universe with a small magnetic field. In cases (1) and (3) electromagnetic radiation 
is produced behind the gravitational wave. In case (2) no electromagnetic radiation appears after the wave 
unless a current is established behind the wave breaking the Maxwell vacuum. In all three cases the 
presence of the magnetic or electric fields results in a modification of the amplitude of the incoming 
gravitational wave which is explicitly calculated using the Einstein--Maxwell vacuum field equations. 
\end{abstract}
\thispagestyle{empty}
\newpage

\section{Introduction}\indent
The existence of large scale magnetic fields in the universe has led to extensive studies of their behavior in 
cosmological models \cite{T1}--\cite{B}. The observation by Marklund, Dunsby and Brodin \cite{M} that gravity 
wave perturbations of Friedmann--Lema\^itre--Robertson--Walker cosmological models encountering weak 
magnetic test fields can produce electromagnetic waves is of particular significance. This phenomenon has 
recently been studied again in cosmology \cite{HOF}. In the present paper we examine it in further detail by 
taking it out of the cosmological setting and by replacing the gravitational waves by a single impulsive wave in 
the following three illustrative situations: (1) a cylindrical impulsive gravitational wave propagating into a 
cylindrically symmetric universe containing an approximately uniform magnetic field (the Bonnor \cite{Bo} universe, 
rediscovered by Melvin \cite{Me}), (2) an axially symmetric impulsive gravitational wave propagating into an 
axially symmetric universe containing an approximately uniform electric field (the Mc Vittie \cite{McV} universe; 
see also \cite{Bon}) and (3) a `spherical' impulsive gravitational wave propagating into a universe with no 
gravitational field but with a weak uniform magnetic test field. In each of these three cases the space--time to the 
future of the null hypersurface history of the impulsive gravitational wave (the model universe left behind by the 
wave) is calculated in a future neighborhood of the null hypersurface, using the Einstein--Maxwell vacuum 
field equations. In cases (1) and (3) we find that electromagnetic radiation is generated behind the gravitational 
wave. In case (2) no electromagnetic radiation appears after the wave unless a current is established behind the 
wave breaking the Maxwell vacuum. In all three cases the presence of the magnetic or electric fields in front 
of the gravitational wave modifies the amplitude of the gravitational wave and this modification is explicitly calculated 
using the Einstein--Maxwell vacuum field equations. The three cases are described in sections 2, 3 and 4 
respectively of this paper followed by a discussion of the main features of our results in section 5. Some useful 
calculations pertinent to section 2 are given in appendix A.

\setcounter{equation}{0}
\section{The Cylindrically Symmetric Case}\indent
The cylindrically symmetric line--element can be written in the form \cite{SKMHH}
\begin{equation}\label{2.1}
ds^2=e^{2k-2U}(dt^2-d\rho ^2)-e^{-2U}\rho ^2 d\phi ^2-e^{2U}dz^2\ ,\end{equation}
where, in general, $k$ and $U$ are functions of $\rho$ and $t$. An example of a static model of a 
gravitational field having a magnetic field as origin is \cite{Bo}, \cite{Me}
\begin{equation}\label{2.2}
e^{2U}=e^{k}=f^2\ ,\qquad f=1+\frac{1}{4}B^2\rho ^2\ ,\end{equation}
with $B$ a real constant. The corresponding Maxwell field is given by the 2--form
\begin{equation}\label{2.3}
F=B\,f^{-2}\rho\,d\rho\wedge d\phi\ .\end{equation}
Referred to an orthonormal tetrad this is a pure magnetic field with one physical component $Bf^{-2}$ 
and thus ``is not a uniform field in the classical sense"\cite{Bo}. For a weak magnetic field, with terms of 
order $B^2$ neglected (more correctly, with dimensionless quantities of order $B^2\rho ^2$ neglected), 
the magnetic field (2.3) is approximately uniform. We wish to have an impulsive gravitational wave propagating 
into this universe. The history of such a wave is a null hypersurface. Respecting the cylindrical symmetry the 
simplest such null hypersurfaces in the space--time with line--element (2.1) have equations
\begin{equation}\label{2.4}
u=t-\rho ={\rm constant}\ .\end{equation}
Such a null hypersurface has the potential to be the history of a cylindrical wave. Changing to $u$ as a coordinate in 
place of $t$ according to (2.4) the line--element (2.1) reads
\begin{equation}\label{2.5}
ds^2=e^{2k-2U}du\,(du+2\,d\rho )-e^{-2U}\rho ^2 d\phi ^2-e^{2U}dz^2\ ,\end{equation}
with $k, U$ functions now of $\rho$ and $u$ in general but given by (2.2) for the magnetic universe above.

To construct a space--time model of a cylindrical impulsive gravitational wave propagating into the magnetic universe, 
with history $u=0$ (say), and leaving behind a cylindrically symmetric Einstein--Maxwell vacuum we proceed as follows: 
We use coordinates labelled $x^\mu =(u, \rho , \phi , z)$ for $\mu =1, 2, 3, 4$. The null hypersurface $\Sigma (u=0)$ divides 
space--time into two halves $M_+(u>0)$ and $M_-(u<0)$. We take $M_-$ to be to the past of $\Sigma$ with line--element 
(2.5) with $U, k$ given by (2.2) and $M_+$ to be to the future of $\Sigma$ with line--element of the form (2.5) and with the 
as yet unknown functions $U, k$ denoted now by $U_+$ and $k_+$. We assume that $\Sigma$ is singular, 
which means that the metric tensor of $M_-\cup M_+$ is only $C^0$ across $\Sigma$ and thus physically $\Sigma$ is in 
general the history of a cylindrically symmetric null shell and/or impulsive gravitational wave (see \cite{BH} for a review 
of singular null hypersurfaces in general relativity). We seek to find the conditions on the functions $U, k, U_+, k_+$ so that 
$u=0$ is the history of an impulsive gravitational wave and not a null shell. The system of coordinates we are using is 
common to the two sides of $\Sigma$. Since the metric tensor is $C^0$ the induced metrics on $\Sigma$ from its embedding 
in $M_+$ and in $M_-$ must be identical and thus we shall have
\begin{equation}\label{2.6}
U_+(u=0, \rho )=\log f\ , \end{equation}
with $f$ given by (2.2). The subset of coordinates $\xi ^a=(\rho , \phi , z)$ with $a=2, 3, 4$, will be taken as intrinsic 
coordinates on $\Sigma$. Here $\rho$ is a parameter running along the generators of $\Sigma$ while $\theta ^A=(\phi , z)$ with 
$A=3, 4$ label the generators. We denote by $e_{(a)}=\partial /\partial\xi ^a$ the tangential basis vectors. Their scalar products 
give the induced metric tensor $g_{ab}$ which is singular since $\Sigma$ is null. The line--element (2.5) restricted to $\Sigma$ 
reads
\begin{equation}\label{2.7}
ds^2|_{\Sigma}=g_{ab}d\xi ^a\,d\xi ^b=e_{(a)}\cdot e_{(b)}d\xi ^a\,d\xi ^b\ .\end{equation}
It is convenient to introduce a pseudo--inverse of $g_{ab}$ (see \cite{BH}) which we denote by $g^{ab}_*$ and which is formed 
by the inverse $g^{AB}$ of $g_{AB}$ bordered by zeros. As normal to $\Sigma$ we take $n^\mu\,\partial /\partial x^\mu =\partial 
/\partial\rho =e_{(2)}$. This vector field is tangent to $\Sigma$ and in order to describe extrinsic properties of $\Sigma$ we introduce 
a transversal vector field $N^\mu$ on $\Sigma$ which for convenience we take to be future--directed, null and orthogonal 
to the two space--like vectors $e_{(A)}$ at each point of $\Sigma$. Thus we have
\begin{equation}\label{2.8}
N_{\mu}\,n^\mu =1\ ,\ N_\mu\,N^\mu =0\ ,\ N_\mu\,e^\mu _{(A)}=0\ .\end{equation}
Thus $N_\mu =(\frac{1}{2}, 1, 0, 0)$. Following the algorithm developed in \cite{BH} we define the transverse extrinsic curvature 
${\cal K}^{\pm}_{ab}$ on either side of $\Sigma$ by
\begin{equation}\label{2.9}
{\cal K}^{\pm}_{ab}=-N_{\mu}(e^\mu _{(a),\lambda}+{}^{\pm}\Gamma ^{\mu}_{\lambda\sigma}\,e^{\sigma}_{(a)})\,e^{\lambda}_{(b)}\ ,
\end{equation}
with the comma denoting partial differentiation with respect to $x^\mu$ and ${}^{\pm}\Gamma ^{\mu}_{\lambda\sigma}$ the components 
of the Riemannian connection calculated on either side of $\Sigma$. The jump in the quantities (2.9) is defined by
\begin{equation}\label{2.10}
\gamma _{ab}=2\,[{\cal K}_{ab}]=2\,({\cal K}^+_{ab}-{\cal K}^-_{ab})\ .\end{equation}
We find in the present case that $\gamma _{ab}=0$ except for
\begin{eqnarray}\label{2.11}
\gamma _{22}&=&-2\,[\Gamma ^2_{22}]\ ,\\
\gamma _{33}&=&-[\Gamma ^1_{33}]-2\,[\Gamma ^2_{33}]\ ,\\
\gamma _{44}&=&-[\Gamma ^1_{44}]-2\,[\Gamma ^2_{44}]\ .\end{eqnarray}
The singular null hypersurface $\Sigma$ can represent the history of a null shell and/or an impulsive gravitational wave. The surface 
stress--energy tensor of the shell, if it exists, is calculated from $\gamma _{ab}$ and is given by (see Eq.(2.37) of \cite{BH})
\begin{equation}\label{2.12}
16\pi\,S^{ab}=\mu\,n^a\,n^b+P\,g^{ab}_*\ ,\end{equation}
with the surface energy density $\mu$ and pressure $P$ defined by
\begin{eqnarray}\label{2.13}
16\pi\,\mu &=&-\gamma _{ab}\,g^{ab}_*=-\gamma _{AB}\,g^{AB}\ ,\\
16\pi\,P&=&-\gamma ^{\dagger}\ ,\end{eqnarray}
In (2.14) $n^a=(1, 0, 0)$ and in (2.16) $\gamma ^{\dagger}=\gamma _{ab}\,n^a\,n^b=\gamma _{22}$. Hence the 
conditions for \emph{no shell} read
\begin{equation}\label{2.14}
\gamma ^{\dagger}=\gamma _{22}=0\ \ {\rm and}\ \ -g^{ab}_*\gamma _{ab}=f^2\gamma _{33}
+f^{-2}\rho ^2\gamma _{44}=0\ .\end{equation}Using (2.11)--(2.13) and calculating the components of the 
Riemannian connection associated with the metric tensor given via the line--element (2.5) we find that the first 
of (2.17) requires that $[k_\rho ]=0$ (with the subscript denoting partial differentiation with respect $\rho$) and the 
second of (2.17) requires that $[k]=0$. Hence the boundary condition for \emph{no shell} is
\begin{equation}\label{2.15}
[k]=0\ ,\end{equation}
expressing the continuity of the function $k(\rho , u)$ across $\Sigma (u=0)$. The gravitational wave part of the signal 
with history $\Sigma$ is described by a part of $\gamma _{ab}$, denoted $\hat\gamma _{ab}$, defined by Eq.(2.47) 
of \cite{BH}:
\begin{equation}\label{2.16}
\hat\gamma _{ab}=\gamma _{ab}-\frac{1}{2}g_{ab}\,g^{cd}_*\gamma _{cd}-2\,\gamma _{(a}N_{b)}+\gamma ^{\dagger}
N_a\,N_b\ ,\end{equation}with $N_a=(1, 0, 0)$. With (2.18) satisfied we find that $\hat\gamma _{ab}=0$ except
\begin{eqnarray}\label{2.17}
\hat\gamma _{33}&=&\gamma _{33}=2\rho ^2f^{-4}[U_u]\ ,\\
\hat\gamma _{44}&=&\gamma _{44}=-2\,[U_u]\ .\end{eqnarray}
The fact that (2.20) and (2.21) are multiples of each other means that the gravitational wave here has only one degree of freedom. Hence we see that \emph{for an impulsive wave} with history 
$\Sigma$ we must have
\begin{equation}\label{2.18}
[U_u]\neq 0\ ,\end{equation}with the subscript denoting partial differentiation with respect to $u$.

The Maxwell field in $M_+\cup M_-$ is given in general by the 2--form
\begin{equation}\label{2.19}
F=w_\rho\,dz\wedge d\rho +w_u\,dz\wedge du +s_\rho\,d\rho\wedge d\phi +s_u\,du\wedge d\phi\ ,\end{equation}
with $w, s$ each functions of $\rho , u$ and the subscripts as always denoting partial derivatives. In $M_-(u<0)$ we 
have
\begin{equation}\label{2.20}
w=0\ ,\qquad s=-2\,B^{-1}f^{-1}\ ,\end{equation}with $f$ given by (2.2). Substitution of (2.24) into (2.23) yields (2.3).

To obtain the space--time $M_+(u\geq 0)$ and the electromagnetic field for $u\geq 0$ we must satisfy the Einstein--Maxwell 
vacuum field equations in $u\geq 0$ with a line--element of the form (2.5) and a Maxwell 2--form of the form (2.23). These 
equations are listed in appendix A. For our purposes it is sufficient to solve these equations for small $u>0$ (i.e. in a future 
neighborhood of $\Sigma$) . The unknown functions of $\rho , u$ are $U, k, w, s$ with $k$ continuous across $u=0$ and $U_u$ 
jumping across $u=0$. Hence for small $u$ we can write
\begin{eqnarray}\label{2.21}
U&=&\log f+u\,\theta (u)\,U_1+O(u^2)\ ,\\
k&=&2\,\log f+u\,\theta (u)\,k_1+O(u^2)\ ,\end{eqnarray}with $f$ given by (2.2).
Here $\theta (u)$ is the Heaviside step function which is equal to unity if $u>0$ and equal to zero if $u<0$. 
For consistency with the expansions (2.25) and (2.26), and in the light of (2.24), 
we also assume that
\begin{eqnarray}\label{2.22}
s&=&-2\,B^{-1}f^{-1}+u\,\theta (u)\,s_1+O(u^2)\ ,\\
w&=&u\,\theta (u)\,w_1+O(u^2)\ .\end{eqnarray}
The unknown functions $U_1, k_1, s_1, w_1$ in (2.25)--(2.28) are functions of $\rho$ only. Substituting (2.25)--(2.28) in Einstein's 
equations (A-3)--(A-7) results in
\begin{equation}\label{2.23}
w_1=0\ ,\end{equation}
and then
\begin{equation}\label{2.24}
\frac{d}{d\rho}(\rho ^{1/2}U_1)= B\,\rho ^{-1/2}s_1\ ,\end{equation}
\begin{equation}\label{2.25}
\frac{1}{\rho}\frac{dk_1}{d\rho}-2\,f^{-1}\frac{df}{d\rho}\,\frac{dU_1}{d\rho}=2\,\frac{B}{\rho}\,\frac{ds_1}{d\rho}+2\,B^2f^{-2}U_1\ ,\end{equation}
\begin{equation}\label{2.26}
\frac{dU_1}{d\rho}-U_1\,f^{-1}\frac{df}{d\rho}+\frac{1}{2\rho}\,U_1=\frac{dk_1}{d\rho}\ ,\end{equation}
\begin{equation}\label{2.27}
2\,U_1f^{-1}\frac{df}{d\rho}-2U^2_1-\frac{1}{\rho}k_1=\frac{2}{\rho ^2}f^2(s_1-B\,\rho\,f^{-2})s_1\ .\end{equation}
Maxwell's equations (A-1) and (A-2) now provide just one extra relevant equation, namely,
\begin{equation}\label{2.28}
\frac{d}{d\rho}(\rho ^{-1/2}f\,s_1)=-B\,f^{-1}\rho ^{1/2}U_1\ .\end{equation}We observe that 
(2.30), (2.31) and (2.34) imply (2.32). Hence the strategy for solving these five equations is to first solve (2.30) and (2.34) for $s_1, U_1$, 
then substitute these solutions into (2.33) to obtain $k_1$ algebraically and then to check that (2.31) is automatically satisfied. Proceeding 
in this way we obtain the solutions 
\begin{eqnarray}\label{2.29}
U_1&=&a_0\,\rho ^{-1/2}f^{-1}\left (1-\frac{1}{4}B^2\rho ^2\right )+b_0\,\rho ^{1/2}B\,f^{-1}\ ,\\
s_1&=&-a_0\,B\rho ^{3/2}f^{-2}+b_0\,\rho ^{1/2}f^{-2}\left (1-\frac{1}{4}B^2\rho ^2\right)\ ,\\
k_1&=&-2\,(a_0^2+b_0^2)-a_0\,B^2\rho ^{3/2}f^{-1}+2\,b_0\,B\rho ^{1/2}f^{-1}\ ,\end{eqnarray}
where $a_0, b_0$ are real constants. A convenient way to interpret these results is to use them to obtain 
information about parts of the Weyl tensor and the Maxwell tensor in $M_+\cup M_-$ on a basis of 1--forms $\theta ^\mu$ ($\mu =1, 2, 3, 4$) in terms of which the 
line--element (2.5) can be written
\begin{equation}\label{2.30}
ds^2=2\,\theta ^1\,\theta ^2-(\theta ^3)^2-(\theta ^4)^2\ .\end{equation}
Such a basis is given by
\begin{equation}\label{2.31}
\theta ^1=e^{2k-2U}(d\rho +\frac{1}{2}du)\ ,\ \theta ^2=du\ ,\ \theta ^3=e^{-U}\rho\,d\phi\ ,\ \theta ^4=e^Udz\ .\end{equation}
The Weyl tensor components on this basis, denoted $C_{\mu\nu\lambda\sigma}$, for the space--time  $M_+\cup M_-$ is 
dominated for small $u$ by the tetrad component
\begin{equation}\label{2.32}
C_{2323}=\{a_0\,\rho ^{-1/2}f^{-1}(1-\frac{1}{4}B^2\rho ^2)+b_0\,\rho ^{1/2}f^{-1}B\}\,\delta (u)\ , \end{equation}
with all other terms and components at most $O(u^0)$. Here $\delta (u)$ is the Dirac delta function. This 
dominant part of the Weyl tensor of  $M_+\cup M_-$ is type N in the Petrov classification with degenerate principal null direction 
$\partial /\partial\rho$. It represents a gravitational wave. When $B=0$ ($f=1$) we see a cylindrical impulsive gravitational wave which is  
singular on the axis $\rho =0$ and having one degree of freedom manifested by the appearance of the real constant $a_0$. The presence 
of the magnetic field $B$ clearly modifies the amplitude of the gravitational wave. We will comment on this modification in section 5 when 
we can compare (2.40) with the examples described in the next two sections. The tetrad components of the Maxwell field in  $M_{\pm}$ 
will be denoted $F^{\pm}_{\mu\nu}$. In general they jump across $\Sigma$ with the jump given by
\begin{equation}\label{2.33}
f_{\mu\nu}=[F_{\mu\nu}]=F^+_{\mu\nu}-F^-_{\mu\nu}\ .\end{equation}In the present case (2.27)--(2.29) and (2.36) mean that $f_{\mu\nu}$ 
vanishes except for
\begin{equation}\label{2.34}
f_{23}=f\,\rho ^{-1}s_1\ .\end{equation}It thus follows that the bivector $f_{\mu\nu}$ is algebraically special 
(type N) in the classification of bivectors with $\partial /\partial\rho$ as degenerate principal null direction.  Thus (2.42) indicates the 
presence of cylindrical electromagnetic waves behind the impulsive gravitational as it propagates through the 
universe with the magnetic field labelled by $B$.

\setcounter{equation}{0}
\section{The Axially Symmetric Case}\indent
We now construct a space--time model of an axially symmetric impulsive gravitational wave propagating into a static axially symmetric 
universe containing an electric field. The line--element of a simple such space--time is  \cite{McV}, \cite{Bon}
\begin{equation}\label{3.1}
ds^2=-W^{-3}dz^2-W^{-1}(d\rho ^2+\rho ^2d\phi ^2)+W\,dt^2\ ,\end{equation}
with $W=(1+E\,z)^2$ and $E$ is a real constant. This is a solution of the Einstein--Maxwell vacuum field equations with the Maxwell 2--form 
given by
\begin{equation}\label{3.2}
F=E\,dz\wedge dt\ .\end{equation}This is clearly a pure electric field. When expressed on an orthonormal tetrad it has the one non--vanishing 
physical component $E\,W$ and so is not a uniform electric field in the classical sense. It is approximately uniform for small dimensionless 
parameter $E\,z$ however. A simple family of null hypersurfaces in this space--time is given by $u={\rm constant}$ with $u$ derived from
\begin{equation}\label{3.3}
du=dt-W^{-2}dz\ .\end{equation}Such null hypersurfaces can act as the histories of the wave--fronts of axially symmetric waves. Using $u$ 
instead of $t$ as a coordinate the line--element (3.1) reads
\begin{equation}\label{3.4}ds^2=W\,du^2+2\,W^{-1}du\,dz-W^{-1}(d\rho ^2+\rho ^2d\phi ^2)\ .\end{equation}Here $z$ is a parameter running 
along the generators of the null hypersurfaces $u={\rm constant}$. It is convenient to work instead with an affine parameter $r$ along 
these generators which is related to $z$ by $1+E\,z=(1-E\,r)^{-1}$. Replacing $z$ by $r$ in (3.4) means the line--element now reads
\begin{equation}\label{3.5}
ds^2=2\,du\,dr+(1-E\,r)^{-2}du^2-(1-E\,r)^2(d\rho ^2+\rho ^2d\phi ^2)\ .\end{equation} The Maxwell field (3.2) now takes the form
\begin{equation}\label{3.6}
F=E\,(1-E\,r)^{-2}dr\wedge du\ .\end{equation}

We now consider a space--time $M_+\cup M_-$ with $M_-(u\leq 0)$ corresponding to the 
space--time with line--element (3.5) having as boundary the null hypersurface $u=0$ and $M_+(u\geq 0)$, with the same boundary $u=0$, 
to be determined. To this latter end we 
solve the vacuum Einstein-Maxwell field equations for $M_+\cup M_-$ requiring that $u=0$ is the history of an axially symmetric impulsive 
gravitational wave (and \emph{not} a null shell). Our objective in doing this is to obtain the space--time $M_+$ with sufficient accuracy to 
determine the coefficient of $\delta (u)$ in the Weyl tensor of $M_+\cup M_-$ and the jump, if it exists, in the Maxwell field across $u=0$, 
in parallel with (2.40)--(2.42). We find that the line--element of $M_+\cup M_-$, for small $u$, can be written in the form (2.38) but with
\begin{eqnarray}\label{3.7}
\theta ^1&=&dr+\frac{1}{2}\left\{(1-E\,r)^{-2}+u\,\theta (u)\,c_1+O(u^2)\right\}\,du\ ,\\
\theta ^2&=&du\ ,\\
\theta ^3&=&(1-E\,r)\{(1+u\,\theta (u)\,\alpha _1+O(u^2))\,d\rho +(u\,\theta (u)\,\beta _1+O(u^2))\,\rho\,d\phi\}\ ,\nonumber\\
\\
\theta ^4&=&(1-E\,r)\{(u\,\theta (u)\,\beta _1+O(u^2))\,d\rho +(1-u\,\theta (u)\,\alpha _1+O(u^2))\,\rho\,d\phi\}\ .\nonumber\\\end{eqnarray}
The functions $\alpha _1, \beta _1, c_1$ are functions of $r, \rho , \phi$. The field equations restrict the functions $\alpha _1$ and $\beta _1$ 
(they also determine the function $c_1$ but we will not require it here) according to
\begin{equation}\label{3.8}
\alpha _1=\frac{\hat\alpha _1(\rho , \phi )}{1-E\,r}\ ,\qquad \beta _1=\frac{\hat\beta _1(\rho , \phi )}{1-E\,r}\ ,\end{equation}
and the functions $\hat\alpha _1,\  \hat\beta _1$ must satisfy the equations
\begin{eqnarray}\label{3.9}
\frac{\partial\hat\alpha _1}{\partial\phi}-\rho\,\frac{\partial\hat\beta _1}{\partial\rho}=2\,\hat\beta _1\ ,\\
\frac{\partial\hat\beta _1}{\partial\phi}+\rho\,\frac{\partial\hat\alpha _1}{\partial\rho}=-2\,\hat\alpha _1\ .\end{eqnarray}
Introducing the complex variable $\zeta =\log\rho +i\phi$ we can integrate (3.12) and (3.13) to arrive at
\begin{equation}\label{3.10}
\hat\alpha _1+i\hat\beta _1=e^{-\bar\zeta}H(\zeta )\ ,\end{equation}
where $H$ is an arbitrary analytic function of $\zeta$. In parallel with (2.40) and (2.42) we find in this case that
\begin{equation}\label{3.11}
C_{2323}-iC_{2324}=\frac{e^{-\bar\zeta}H(\zeta )}{(1-E\,r)}\,\delta (u)\ ,\end{equation}
and
\begin{equation}\label{3.12}
f_{\mu\nu}=[F_{\mu\nu}]=F^+_{\mu\nu}-F^-_{\mu\nu}=0\ .\end{equation}In (3.15) we see an axially symmetric impulsive 
gravitational wave propagating into the universe with the electric field labelled by the parameter $E$. We also see that the 
presence of the electric field modifies the amplitude of the wave by the appearance of $E$ in the coefficient of the delta function. 
The coefficient of the delta function in the Weyl tensor is type N in the Petrov classification with $\partial /\partial r$ in this case 
as degenerate principal null direction (propagation direction in space--time).  
On account of (3.16) there is no electromagnetic radiation immediately behind the gravitational wave in this case. If we were 
to relax the Maxwell vacuum conditions in $M_+$ we can obtain a 4--current with tetrad components
\begin{eqnarray}\label{3.13}
J_3&=&\frac{2\,E}{1-E\,r}\,f_{32}+O(u)\ ,\\
J_4&=&\frac{2\,E}{1-E\,r}\,f_{42}+O(u)\ .\end{eqnarray}A bivector $f_{\mu\nu}$ having only $f_{32}$ and $f_{42}$ non--zero 
is of radiative type with propagation direction $\partial /\partial r$ and represents electromagnetic radiation. 

\setcounter{equation}{0}
\section{The `Spherically' Symmetric Case}\indent
Starting with the line--element of Minkowskian space--time in rectangular Cartesian coordinates and time $X, Y, Z, T$ which 
reads
\begin{equation}\label{4.1}
ds^2=-dX^2-dY^2-dZ^2+dT^2\ ,\end{equation}we make the coordinate transformation
\begin{equation}\label{4.2}
X+iY=r\,G^{1/2}\,e^{iy}\ ,\ Z=r\,x\ ,\ T=u+r\ ,\end{equation}with $G=1-x^2$ then (4.1) takes the form
\begin{equation}\label{4.3}
ds^2=2\,du\,dr+du^2-r^2(G^{-1}dx^2+G\,dy^2)\ .\end{equation}Here $u={\rm constant}$ are future null cones with 
vertices on the time--like geodesic $r=0$, $r$ is an affine parameter along the generators of the null cones and the generators 
are labelled by $x, y$. Using (4.2) again we see that
\begin{equation}\label{4.4}
dX\wedge dY=r\,G\,dr\wedge dy-r^2x\,dx\wedge dy\ .\end{equation}Thus in particular the Maxwell 2--form
\begin{equation}\label{4.5}
F=B\,r\,G\,dr\wedge dy-B\,r^2x\,dx\wedge dy\ ,\end{equation} with $B$ are real constant is a uniform magnetic field. We shall 
restrict considerations to a weak magnetic field in which squares and higher powers of $B$ will be neglected. In this case 
Minkowskian space--time with the bivector (4.5) constitute an approximate solution of the Einstein--Maxwell vacuum field 
equations. To construct a model of a `spherical' impulsive gravitational wave propagating into this universe we will take for 
$M_-(u\leq 0)$ the space--time with line--element (4.3) and the future null cone $u=0$ for the history of the wave. Since the 
future null cone is the history of a 2--sphere expanding with the speed of light we will refer to the gravitational wave with 
history $u=0$ as a `spherical' wave. The reason for the inverted commas is because such a wave will be found to have 
singular points on its spherical wave front, thus violating strict spherical symmetry (see below). Something like this 
is to be expected in general relativity on account of the Birkhoff theorem (see \cite{BH} section 1.2). Now the line--element 
of $M_+\cup M_-$, for small $u$, can be written in the form (2.38) but with
\begin{eqnarray}\label{4.6}
\theta ^1&=&dr+\frac{1}{2}\left\{1+u\,\theta (u)\,c_1+O(u^2)\right\}\,du\ ,\\
\theta ^2&=&du\ ,\\
\theta ^3&=&r\,G^{-1/2}(1+u\,\theta (u)\,\alpha _1+O(u^2))\,dx +r\,G^{1/2}(u\,\theta (u)\,\beta _1+O(u^2))\,dy\ ,\nonumber\\
\\
\theta ^4&=&r\,G^{-1/2}(u\,\theta (u)\,\beta _1+O(u^2))\,dx +r\,G^{1/2}(1-u\,\theta (u)\,\alpha _1+O(u^2))\,dy\ .\nonumber\\\end{eqnarray}
Here the functions $\alpha _1, \beta _1,\ c_1$, along with the functions $f_{\mu\nu}$, are functions for $x, y, r$ and can be determined from the vacuum Einstein--Maxwell 
field equations (in particular ensuring by the vacuum conditions that no null shell can have $u=0$ as history and also that there is 
no surface 4--current on $u=0$). As in the example of section 3 we shall not require the function $c_1$ although it can be determined 
using the field equations of course. From Maxwell's equations we find that
\begin{eqnarray}\label{4.7}
\frac{\partial}{\partial r}(r\,f_{32})=-B\,r\,G^{1/2}\beta _1\ ,\\
\frac{\partial}{\partial r}(r\,f_{42})=B\,r\,G^{1/2}\alpha _1\ .\end{eqnarray}
Neglecting $O(B^2r^2)$--terms we conclude that
\begin{equation}\label{4.8}
B\,r\,G^{1/2}f_{32}=K(x, y)\ \qquad {\rm and}\qquad B\,r\,G^{1/2}f_{42}=L(x, y)\ ,\end{equation} with $K$ and $L$ arbitrary functions of $x, y$. Einstein's equations 
with the electromagnetic energy--momentum tensor as source yield
\begin{eqnarray}\label{4.9}
\frac{\partial}{\partial r}(r\,\alpha _1)&=&B\,r\,G^{1/2}f_{42}\ ,\\
\frac{\partial}{\partial r}(r\,\beta _1)&=&-B\,r\,G^{1/2}f_{32}\ ,\end{eqnarray}
from which we conclude that, in the light of (4.12),
\begin{equation}\label{4.10}
\alpha _1=L(x,y) +\frac{C(x, y)}{r}\ ,\qquad \beta _1=-K(x,y) +\frac{D(x, y)}{r}\ ,\end{equation} where $C$ and $D$ are arbitrary functions of $x, y$. Now defining
\begin{equation}\label{4.11}
\zeta =\frac{1}{2}\log\left (\frac{1+x}{1-x}\right )+iy\ ,\end{equation}the remaining Einstein field equations give the 
single complex equation
\begin{equation}\label{4.12}
\frac{\partial}{\partial\zeta}\{G\,(\alpha _1+i\beta _1)\}=-G\,x\,(L-iK)\ ,\end{equation}from which we conclude, using (4.15), that
\begin{equation}\label{4.13}
G\,(C+iD)=\bar {\cal F}(\bar\zeta)\qquad {\rm and}\qquad L-iK=\bar {\cal G}(\bar\zeta)\ ,\end{equation}where ${\cal F}, {\cal G}$ are 
arbitrary analytic functions. Thus (4.12) and (4.15) now read
\begin{equation}\label{4.14}
f_{42}+if_{32}=B^{-1}G^{-1/2}\frac{1}{r}\,{\cal G}(\zeta )\ ,\end{equation}and
\begin{equation}\label{4.15}
\alpha _1-i\beta _1=\frac{1}{r}G^{-1}{\cal F}(\zeta )+{\cal G}(\zeta )\ ,\end{equation}respectively.

In this case the delta function part of the Weyl tensor is given by 
\begin{equation}\label{4.16}
C_{2323}-iC_{2324}=(\alpha _1-i\beta _1)\,\delta (u)=\left\{\frac{1}{r}G^{-1}{\cal F}(\zeta )+{\cal G}(\zeta )\right\}\,\delta (u)\ .\end{equation}
The first term in the coefficient of the delta function is the amplitude of a `spherical' wave with the expected ``directional" 
singularities at $x=\pm 1$ (corresponding to $G(x)=0$, equivalently $X=Y=0$) while the second term is the modification to the amplitude due to the wave 
encountering the weak magnetic field. In (4.19) we see the algebraically special jumps in the Maxwell field across $u=0$ 
which indicate the presence of electromagnetic radiation in the region $M_+$ of space--time to the future of the history of the 
impulsive gravitational wave (i.e. behind the wave).This radiation is spherical--fronted ($u={\rm contant}>0$ being the histories 
of the wave--fronts), singular at $r=0$ and also has directional singularities at $x=\pm 1$.

\setcounter{equation}{0}
\section{Discussion}\indent
In sections 2, 3 and 4 above we have considered an impulsive gravitational wave propagating into a vacuum universe with a 
magnetic field present in the first and last cases and into a vacuum universe with an electric field present in the second case. If the 
vacuum is preserved after the wave has passed then in the region behind the wave electromagnetic radiation appears in the 
first and last cases but not in the second case. In addition we have found that in all three cases the amplitude of the impulsive 
gravitational wave is modified by the existence of the magnetic or electric field that it encounters. There is an interesting 
pattern to this modification when the magnetic and electric fields are weak in all three cases. In the cylindrically symmetric case, 
combining (2.40) and (2.42) with (2.36), we can write, approximately for small $B$,
\begin{equation}\label{5.1}
C_{2323}=\{a_0\rho ^{-1/2}+\rho\,B\,f_{23}\}\,\delta (u)\ ,\end{equation}
for the delta function part of the Weyl tensor. The coefficient of the delta function here is a sum of a cylindrical wave term and 
an interaction between the weak magnetic field and the electromagnetic radiation. For the axially symmetric case with a weak 
electric field we have from (3.15)
\begin{equation}\label{5.2}
C_{2323}-iC_{2324}=\{e^{-\bar\zeta}H(\zeta )+r\,E\,e^{-\bar\zeta}H(\zeta ) \}\,\delta (u)\ .\end{equation}
In this case there is no electromagnetic radiation generated behind the gravitational wave but the coefficient of the delta function 
is the sum of an axially symmetric wave term and an interaction between the weak electric field and the gravitational radiation. Finally 
in the `spherical' case we have (4.21) which with (4.19) can be written
\begin{equation}\label{5.3}
C_{2323}-iC_{2324}=\left\{\frac{1}{r}G^{-1}{\cal F}(\zeta )+r\,B\,G^{1/2}(f_{42}+if_{32})\right\}\,\delta (u)\ .\end{equation}
The coefficient of the delta function here is a sum of a `spherical'  wave and an interaction between the weak magnetic 
field and the electromagnetic radiation.

When electromagnetic radiation appears above it takes the form of an electromagnetic shock wave accompanying the 
impulsive gravitational wave. The history of the electromagnetic shock wave is the null hypersurface $u=0$. Should 
we wish to know the field in $u>0$, to the future of the history of the wave, we would require, for example, the 
$O(u^2)$--terms in (2.25)--(2.28), (3.7)--(3.10) and (4.6)--(4.9). The examples given in this paper have motivated the development of a general, 
relativistically invariant, treatment of the interaction of impulsive gravitational waves with electromagnetic fields which 
will be described in a future publication.

\appendix
\section{Cylindrically Symmetric Einstein--Maxwell Equations } \setcounter{equation}{0}
With a metric tensor given by the line--element (2.5) and a Maxwell tensor given by the 2--form (2.23) the 
Maxwell field equations ($d{}^*F=0$ with ${}^*F$ the dual of $F$) are given by (with subscripts denoting partial 
differentiation)\begin{eqnarray}\label{A1}
\rho\,(e^{-2U}w_\rho )_u+(\rho\,e^{-2U}(w_u-w_\rho ))_\rho &=&0\ ,\\
\rho ^{-1}(e^{2U}s_\rho )_u+(\rho ^{-1}e^{2U}(s_u-s_\rho ))_\rho &=&0\ .
\end{eqnarray}Einstein's equations ($R_{\mu\nu}=-2\,E_{\mu\nu}$ with $R_{\mu\nu}$ the Ricci tensor and $E_{\mu\nu}$ 
the electromagnetic energy--momentum tensor) read
\begin{equation}\label{A2}
k_{\rho\rho}-2\,k_{u\rho}+\frac{1}{\rho}(k_\rho -k_u)-U_{\rho\rho}+2\,U_{u\rho}-\frac{1}{\rho}(U_\rho -U_u)-2\,U_u^2
=T_1\ ,\end{equation}
\begin{equation}\label{A3}
2\,U_{u\rho}-U_{\rho\rho}+\frac{1}{\rho}(U_u-U_\rho)-2\,U_u\,U_\rho +k_{\rho\rho}-2\,k_{u\rho}+\frac{1}{\rho}k_\rho 
=T_2\ ,\end{equation}
\begin{equation}\label{A4}
-U_\rho -\rho\,U_{\rho\rho}+U_u+2\,\rho\,U_{u\rho}=T_3\ ,\end{equation}
and
\begin{equation}\label{A5}
w_\rho\,s_\rho =s_u\,w_\rho +s_\rho\,w_u\ ,\end{equation}
with
\begin{eqnarray}\label{A6}
T_1&=&\frac{2}{\rho ^2}e^{2U}(s_u^2-s_u\,s_\rho +s^2_\rho )
+2\,e^{-2U}(w^2_u-w_uw_\rho +\frac{1}{2}w^2_\rho )\ ,\\
T_2&=&\frac{1}{\rho ^2}e^{2U}s^2_\rho +e^{-2U}w^2_\rho\ ,\\                  
T_3&=&\frac{2}{\rho}e^{-2U}\left \{s_\rho\,s_u-\frac{1}{2}s^2_\rho-\rho ^2(w_\rho\,w_u-\frac{1}{2}w^2_\rho )\right \}\ .\end{eqnarray}


\begin{thebibliography}{99}
\bibitem{T1} C. G. Tsagas, Classical Quantum Gravity {\bf 22}, 393 (2005).
\bibitem{T2} C. G. Tsagas and A. Kandus, Phys. Rev. D{\bf 71}, 123506 (2005).
\bibitem{T3} C. G. Tsagas, Phys. Rev. D{\bf 72}, 123509 (2005).
\bibitem{T4} C. G. Tsagas, Phys. Rev. D{\bf 75}, 087901 (2007).
\bibitem{B} J. D. Barrow, R. Maartens and C. G. Tsagas, Phys. Rep. {\bf 449}, 131 (2007).
\bibitem{M} M. Marklund, P. K. S. Dunsby and G. Brodin, Phys. Rev. D{\bf 62}, 101501(R) (2000).
\bibitem{HOF} P. A. Hogan and S. O'Farrell, Phys. Rev. D{\bf 79}, 104028 (2009).
\bibitem{Bo} W. B. Bonnor, Proc. Phys. Soc. A{\bf 67}, 225 (1954).
\bibitem{Me} M. A. Melvin, Phys. Lett. {\bf 8}, 65 (1964).
\bibitem{McV} G. C. Mc Vittie, Proc. Roy. Soc. A{\bf 124}, 366 (1929).
\bibitem{Bon} W. B. Bonnor, Proc. Phys. Soc. A{\bf 66}, 145 (1953).
\bibitem{SKMHH} H. Stephani, D. Kramer, M. MacCallum, C. Hoenselaers and E. Herlt, ``Exact Solutions to Einstein's 
Field Equations", 2nd edition (Cambridge University Press, 2003).
\bibitem{BH} C. Barrab\`es and P. A. Hogan, ``Singular Null Hypersurfaces in General Relativity" (World Scientific, 2004).
\end{thebibliography}
\end{document}